# Modelling textile structures using bicontinuous surfaces

Chelsea E. Knittel[a,b], Michael Tanis[c], Amy L. Stoltzfus[b], Toen Castle[c,d], Randall D. Kamien[c], Genevieve Dion[b,d].

[a]College of Engineering, Drexel University, Philadelphia, United States of America
[b]The Center for Functional Fabrics, Drexel University, Philadelphia, United States of America
[c]Department of Physics and Astronomy, University of Pennsylvania, Philadelphia, United States of America
[d]Department of Mathematics and Statistics, La Trobe University, Australia
[e]Westphal College of Media, Arts, and Design, Drexel University, Philadelphia, United States

We present a method for modelling textile structures, such as weft knits, on families of bicontinuous surfaces. By developing a tangible interpretation of mathematical theory, we combine perspectives of art, design, engineering, and science to understand how the architecture of the knit relates to its physical and mathematical properties. While modelling and design tools have become ubiquitous in many industries, there is still a significant lack of predictive advanced manufacturing techniques available for the design and manufacture of textiles. We describe a mathematical structure as a system for dynamic modelling of textiles in the form of a physical prototype which may be used to inform and predict relevant textile parameters prior to fabrication. This includes dimensional changes due to yarn relaxation, which would streamline production of knit textiles for industry, makers and textile artists.

*Corresponding author. Email: gd63@drexel.edu

1. Introduction

Automatic textile manufacturing techniques such as knitting and weaving have been established for centuries. Though these technologies have progressed significantly since their initial development, the accompanying modelling and design tools have not reached the level of capability available for other manufacturing techniques such as computer numerical control (CNC) machines and 3D printing. There exists little support for accurate and rapid prototyping of fabrics, both in traditional hand crafting and making, and in novel applications, such as smart textile research and development. This leads to wasted materials and time, as products must be designed largely through trial and error. To address this, we have developed a physical prototype representing a platform for the development of a parameterised model for textile structures to help overcome current barriers to innovation in the textile field. This work describes an adaptive geometrical mesh which provides a topological framework to inform material pathways for modelling textiles made with any material in the form of strings, strands, threads, or yarns. This physical representation suggests a virtual system for dynamic and predictive modelling of textiles based on families of bicontinuous surfaces which automatically incorporate yarn pathways and topologies, allowing for local variations in geometry, including any intermeshing and inter-looping techniques that may be used for manufacturing fabrics by machine or by hand.

1.1 *Modelling of Textiles*

The question of how to model and predict textile structures is not new and has been explored by numerous authors who sought to find methods of predicting the physical appearance and properties of knit and woven structures. Early explorations began in the 1930s with the work of authors such as Frederick Peirce (1937), whose work developed mathematical descriptions of fabric structures to predict and control their properties, as well as other early authors such as Hotte (1950) and Leaf and Anandjiwala (1985), who both pursued study of the relationship between fabric structure and physical properties. These authors laid the ground work for further studies on fabric modelling (Liu et al. 2017) (Poincloux, Adda-Bedia, and Lechenault 2018). For comprehensive reviews of the numerous works studying textile structure and modelling, the authors refer the reader to review articles by Hu, *et al.* (2009), Long, Burns and Yang (2011), and Jevsnik, *et al.* (2014) which cover some of the many techniques that have been explored.

Recently, notable advancements were made towards parametric modelling for woven structures by Guest, *et al.* These authors created a system for topology optimisation of microstructure materials, including 3D woven lattices (Zhao et al. 2014) (Zhang et al. 2015)(Zhao et al. 2016) . Using the "Heaviside Projection Method" described in their earlier work for 3D woven materials, they optimised structure topologies with consideration to manufacturability (Guest, Prévost, and Belytschko 2004). Using these functions, they predicted the permeability of woven structures, optimising performance, with minimal effect on other material properties. Along these lines, they describe an envisioned parameterised system for design using these topology optimisation principles which define the objective function, while taking into consideration manufacturing parameters and mechanical property constraints (Osanov and Guest 2016). As Guest, *et al.* have made notable contributions towards topological optimisation of woven structures, in this study we focus on the novelty of our method as applied to the modelling of knitted loops, although our methodology could also be applied to woven fabrics.

An alternative approach explores the topology of periodic woven and knit structures using a geometric scaffold and the ideas of hyperbolic geometry (Evans, Robins, and Hyde 2013). In this approach, triply periodic minimal surfaces are used as a substrate on which the strands travel as they weave, knit, pack or intertwine together - the method is ambivalent about the nature of the structure. Due to the periodicity in three dimensions, in general the model creates bulk strand structures, however with appropriate choice of parameters, this bulk can be composed of adjacent layers, knit from any periodically repeating stitch. The method we introduce here is related to this approach, but we vary the topology of the underlying substrate to model combinations of different stitch types.

Conventional topological approaches have also been explored for the knit structure. For example, authors Grishanov, *et al.* investigated technology independent classification of textile structures using knot theory (Grishanov, Meshkov, and Omelchenko 2009a, 2009b). They developed a system of classifying multiple fabric structures including weft and warp knits, in addition to woven structure. While they determined that integer invariants (*e.g.*, crossing number, linking number, *etc.*) could not be used to distinguish structurally different textiles, the authors proposed a new "Kaufmann-type" polynomial invariant which they utilised for classification of textile structure which are topologically different. (Grishanov, Meshkov, and Omelchenko 2009b). As it is the geometry, not the topology that varies after the fabric is knit, this work complements our approach.

In an effort to streamline design of knit structure, Narayanan et al. have explored modelling of knit structures with automatic generation of knitting machine instructions likening it to a system as "easy" to use as 3D printing (Narayanan et al. 2018). By inputting a 3d mesh of the desired shape, algorithmic methods are used to convert these forms into knit patterns, compatible with any machine knitting software. Currently this method represents a notable step towards parametric modelling of knit structures by predicting the patterns for complex 3-dimensional knit structures but does not yet give indication of other physical properties such as stretch. Additionally, while scale is predictable using these methods, it requires measurement of a gauge swatch to calculate sizing.

Finally, in the industrial sector, some modelling software exist which provide accurate virtual representations of textile structures. These include the Shima Seiki SDS One Apex software and Stoll Pattern Software M1 Plus for weft knit, Texion Software Solutions ProCad Professional Raschel Office for warp knits, and programs such as PixeLoom, Fiberworks PCW and Weave Point for woven fabrics. The Shima Seiki software suite also provides simulation capabilities for woven fabrics. Yet, there still does not exist a parametric, physics-based software for the design of textile structures. While all of these software packages can provide accurate representations of the visual appearance of textile structures, they lack the ability to account for the effects of yarn material used or yarn relaxation effects which cause changes in the dimension of a fabric after manufacture. We now present a new tool that we believe will facilitate the development of this effort.

2. **The weft knit structure**

While our model could be used with any strand form (structure based) textile, here we primarily focus on the weft knit structure to illustrate the use of our system for modelling using bicontinuous surfaces. Knitting is a method of producing a fabric by creating series of intermeshing loops of yarn. While weft knitting can be carried out both by hand and by machines with varying levels of automation, structurally it is always produced from the same base unit, the knit *stitch*, or *loop*. The knit stitch is formed when yarn is drawn through a previously existing loop, and this process is repeated stitch by stitch, row by row, building up the fabric. With a single bed machine, only knit stitches can be produced, formed when the yarn is drawn through the loop from back to front. A double bed machine enables other stitch types to also be produced on the same textile plane. This includes the purl stitch, formed when the yarn is drawn through the loop from front to back. It is important to observe that the knit and purl are the reverse of one another and therefore structurally symmetric: the side from which they are viewed determines their nomenclature. The knit can be identified by the visual cue of small "v" shapes created by the legs of the loops, as shown in figure 1A. The purl can be identified by the small curves of the heads and tails of the loop, as shown in figure 1B.

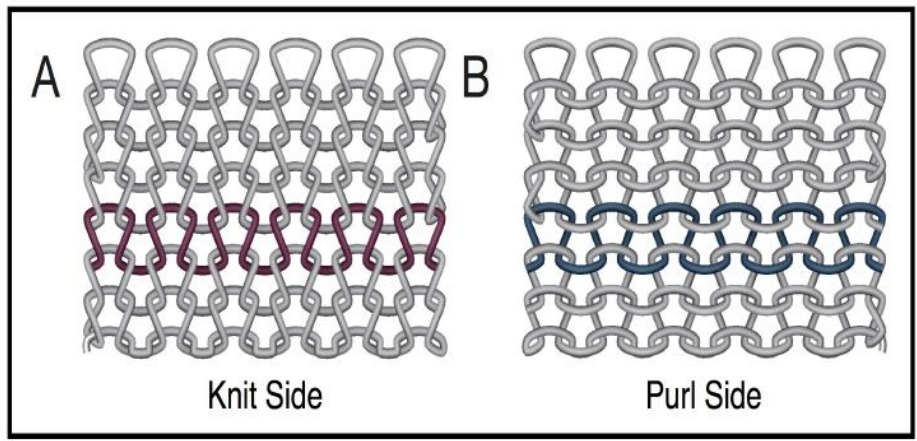

Figure 1. Schematic of the weft knit loop structure. From the front, A) knit (K), and from the back B) purl (P).

During the weft knitting process, yarn loops are held in tension on the needles. However, as soon as the loops leave the needles, this tension is released, and the yarns begin to relax. This relaxation can cause significant changes in the textiles behaviour, most notably, the characteristic rolling behaviours which occur at the edges of the fabric, as well as changes in the length and width. When a fabric of all knit or all purl stitches is produced, rolling will occur at all edges of the fabric. This rolling always occurs towards the knit side of the fabric at the top and bottom edges and towards the purl side of the fabric at the side edges, as shown in an actual knit textile in figure 2. As we will argue, our model explains this behaviour.

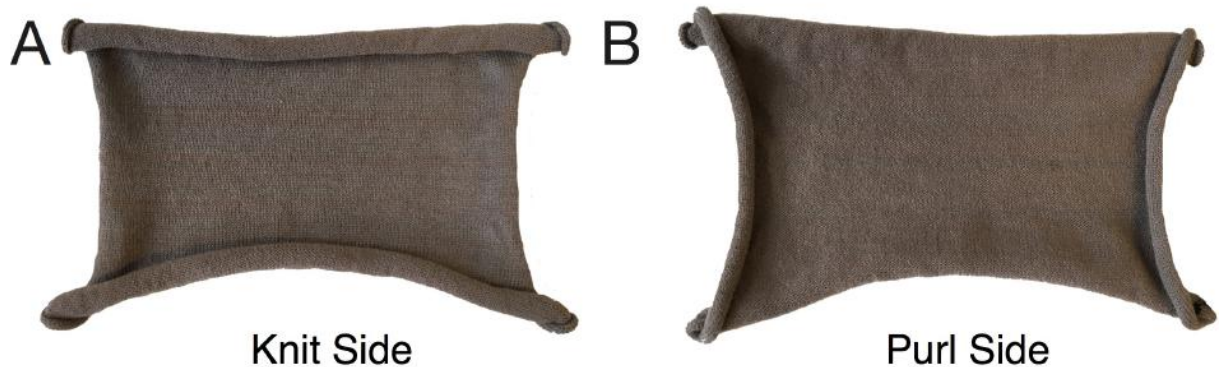

Figure 2. All knit fabric structure; A) front and, B) back, demonstrating characteristic edge rolling behaviours.

The key to the construction of our model is to use a surface which divides space into two distinct regions ("bicontinuous") and is, in some sense, the negative space of the fabric. The surface completely encodes the topology of fabric and so local deformations do not change the interlocking of fibres, only their relative geometry.

3. **Bicontinuous surfaces as a textile lattice**

The physical prototype of the proposed modelling system is depicted in figure 3. The scaffolding consists of a bicontinuous surface which provides a scaffolding on which yarns may lay. The physical model consists of a checkerboard lattice of helicoids, alternating between left- and right-handed helicoids, with the working surface taken along a diagonal cross section of the mathematical lattice shown in figure 4a. Figure 4b depicts the unit cell of the lattice for a weft knit structure. While in the physical model the periodicity of these helicoids is fixed, in a virtual model the periodicity as well as the spacing and width of the helicoids could easily be altered to correspond to various machine gauges and yarn diameters.

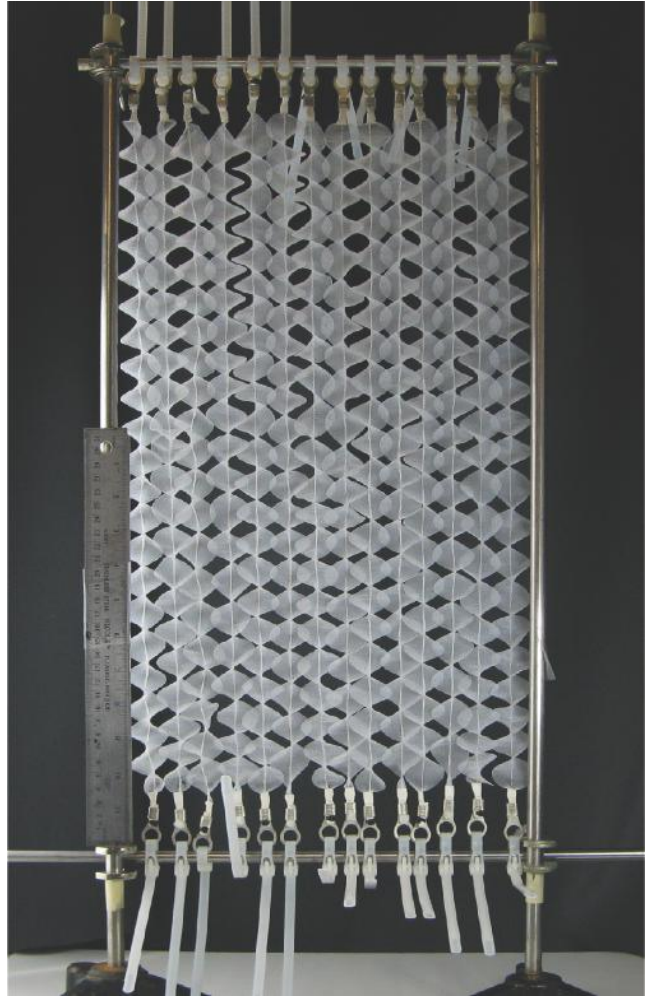

Figure 3. The physical prototype of the bicontinuous helicoid lattice.

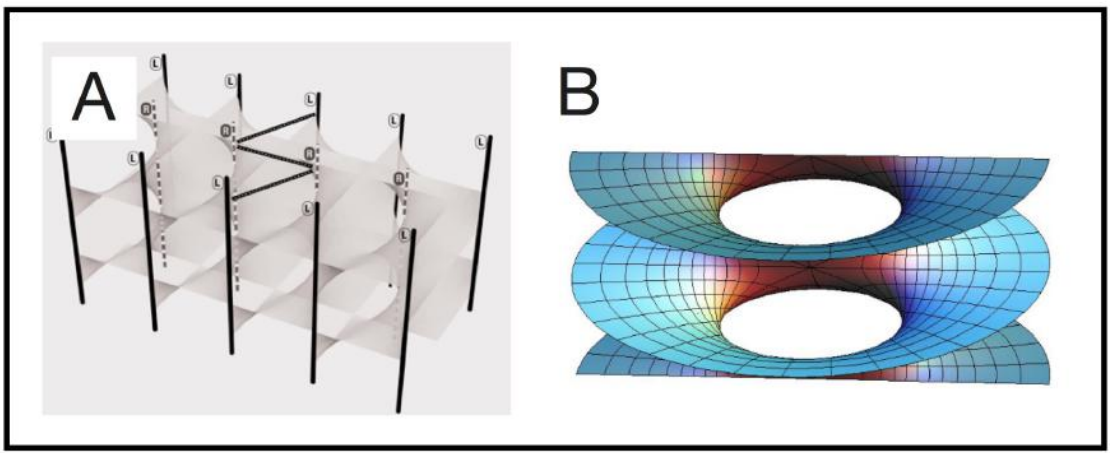

Figure 4. A) A virtual rendition of the bicontinuous helicoid lattice, B) the unit cell of the helicoid lattice. (Generated in 3d-xplormath.org (Palais, n.d.))

The advantages of this method lie in its ability to define and design bicontinuous surfaces additively (Santangelo and Kamien 2006), (Santangelo and Kamien 2007), (Matsumoto, Kamien, and Santangelo 2012) which facilitates mathematical modelling of the scaffolding which can subsequently be input into developed software. To illustrate this concept, we consider the weft knit structure, shown on the helicoid lattice in figure 5, which uses the stitches knit (K), and in figure 6, which uses the stitches knit (K) and purl (P). In weft knit designs, the knit pattern defines and is described by the topology of the yarn. As this topology is fixed (the strand cannot pass through itself or other strands) we can envisage a separating surface between two pieces of touching yarn: each piece must stay on its own side. If we extend these patches of surface away from the contact areas in a consistent and logical way we create a single bicontinuous surface which divides space into two labyrinthine networks. Any piece of yarn is limited to just one channel, but it can still interact topologically with other yarns in the other channel by wrapping around them as they each travel through their own network.

Having created a substrate with bespoke topology, we could, at this point optimise the geometry of the yarns on these surfaces by searching a low-dimensional parameter space. This would allow for prediction of the resulting relaxed knit geometry. However, we leave that for future work and focus here on the knit network structure.

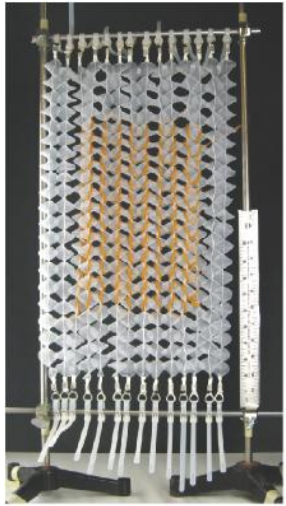 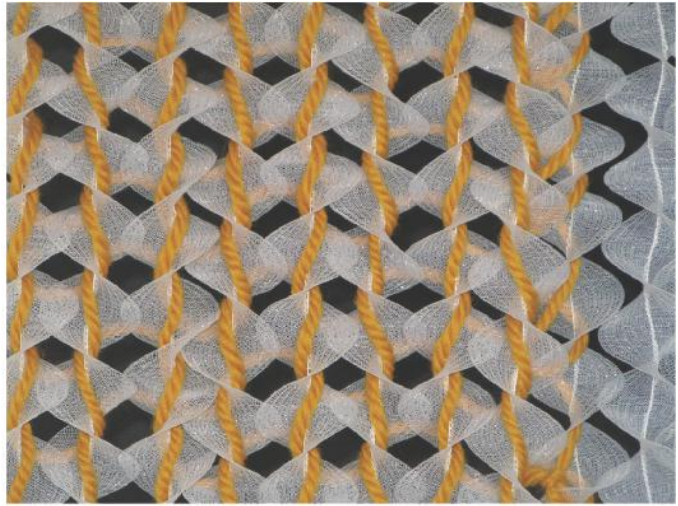
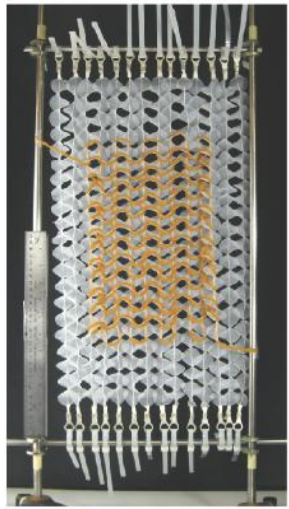 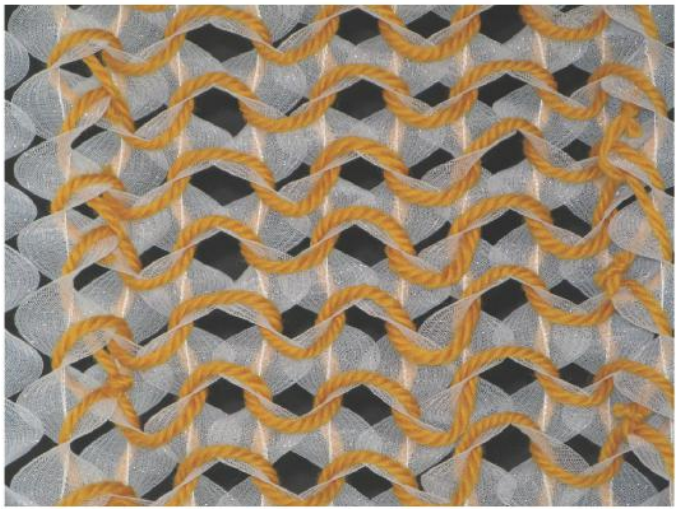

Figure 5. All knit structure on the helicoid lattice; A) front view, B) close up of front view, C) back view, D) close up of back view.

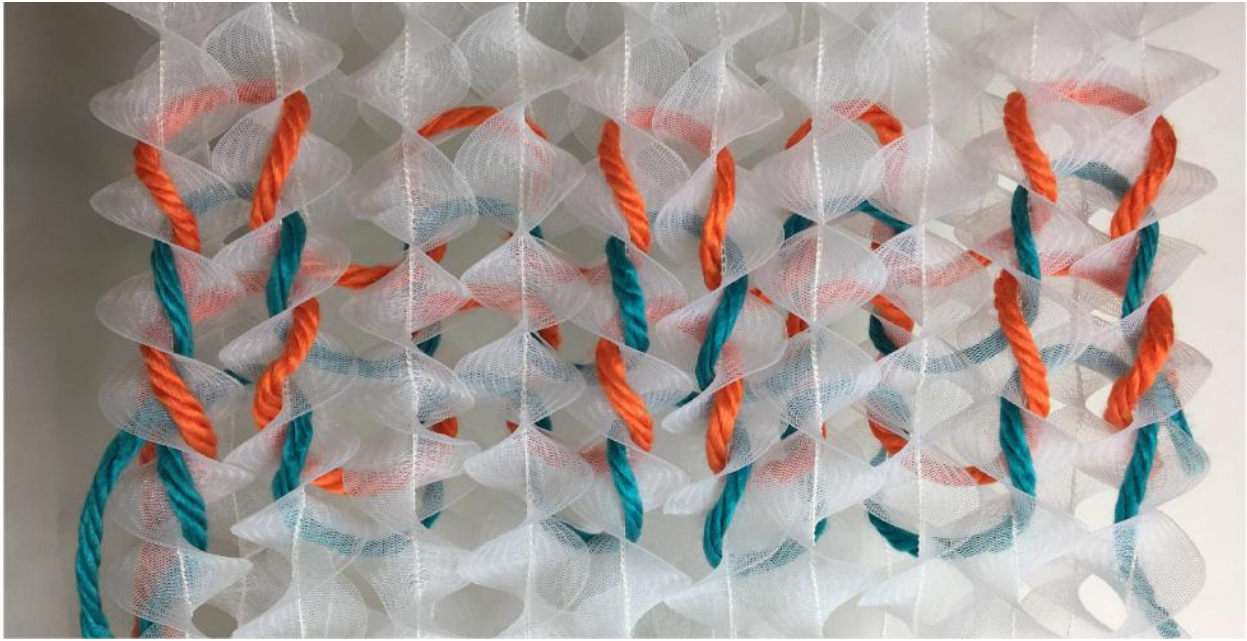

Figure 6. Rib knit structure on the double layer helicoid lattice.

First, we note the handedness to the way that one strand of yarn goes around the other as the stitches are formed along the course direction. Considered from left to right, a K is constructed from a left twist followed by a right twist, while P is a constructed from a right twist followed by left twist. From this it follows that when the fabric is turned over, or viewed from the back, a K becomes a P, and P becomes a K. To build up a sequence of left and right twists, we can generate a three-dimensional surface constructed from a two-dimensional array of helicoids, either left(L)- or right(R)- handed. Reading from left to right, we can then replace a sequence of Ks and Ps with LR and RL, respectively. For example, the knit pattern "KPPKPP…" (knit one, purl two) becomes "LRRLRLLRRLRL…" A two dimensional "checkerboard" arrangement of L and R allows us to knit arbitrary structures: switching from K to P requires us to move from one row of the checkerboard to the other, in order to put two Rs or two Ls in a row. Fortunately, helicoids have a natural home in two-dimensions: they are the Riemann surfaces of the complex natural logarithm of the complex variable

$$z = x + iy, \qquad (1)$$

$$\Phi_R(z;z_o) = \ln(z-z_o) \qquad (2)$$

for R helicoids and

$$\Phi_L(z;z_o) = -\ln(z-z_o) \tag{3}$$

for L helicoids centred at

$$z_o = x_o + iy_o. \tag{4}$$

In order to construct a general surface all that is necessary is to sum an arbitrary combination of $N$ Rs and Ls we write the surface as a graph (*i.e.*, in Monge gauge):

$$h(x,y) = \text{Im}\left[\sum_{n=1}^{N} b_n \Phi_{H_n}(x + iy; z_n)\right] \tag{5}$$

where the $n^{th}$ helicoid has handedness $H_n$ =L or R, is located at

$$z_n = x_n + iy_n, \tag{6}$$

and has magnitude $b_n > 0$. Im gives the imaginary part of the complex function. Thus, all together, we can adjust the positions ($x_n$, $y_n$) and the periodicity (through $b_n$) of all the helicoids. The magnitude controls the periodicity of the stitch. This technique has been used to construct a myriad of complex, bicontinuous surfaces, including "Schnerk's first surface," a surface which is made from a checkerboard array of L and R helicoids. Knitting can then proceed by staying on the first row of LRLRLR until purling is desired, at which point the stitching moves one row back to access RLRLRL and so on. This moving from one row to the other can be done at any point along the course (row) of knitting.

In doing so, we will obtain the three-dimensional path of the yarns which can be used to calculate both bending and stretching. Varying over the parameters allows minimisation of the total yarn energy and prediction of the finished geometry, which will result in an accurate virtual representation of the desired textile which considers boundary conditions, fabrication processes, and yarn size. Just as a second needle bed is required to incorporate knit and purl stitches within the same textile, a second layer of helicoids is necessary to model links-links and other multi-layer knit structures.

### 4. Predicting textile relaxation and deformation behaviours

The following series of figures demonstrate how our model can be used to predict an important textile manufacturing effect; fabric deformation due to yarn relaxation. Figure 7 shows the structure of the plain weft knit and demonstrates how our model informs the structure deformation after the loops are released from tension and the fabric is allowed to relax, simulating the effect that occurs as the fabric is released from the knitting needles during manufacture. Shown from left to right is the fabric in tension viewed from the front, and back, and in its relaxed state viewed from the front, and back. With a release of tension, it can be seen that a) the fabric length shrinks, as the loops relax upward and b) that the fabric width decreases. The apparent decrease in width of the fabric is due to the rolling behaviour which is more easily observed from the back of the fabric. The top and bottom edge rolling is not observable in the physical model as the model is fixed at the top edge and, is weighed down by gravity at the bottom edge.

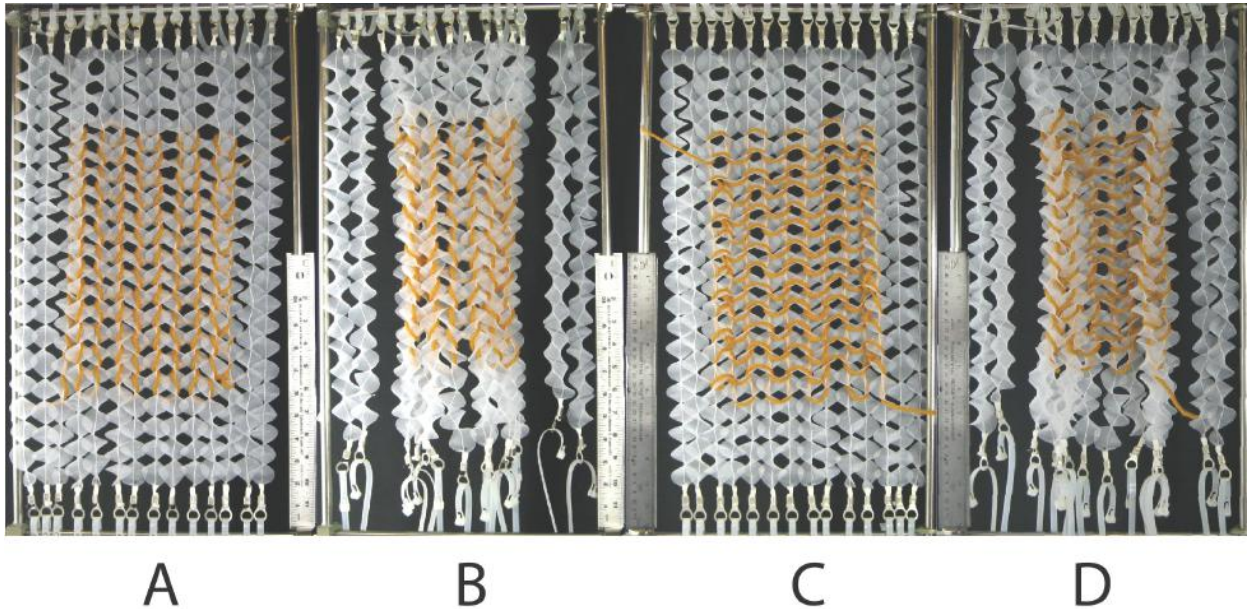

Figure 7. All knit structure on the helicoid lattice; A) front view in tension, B) front view relaxed, C) back view in tension, D) back view relaxed.

Looking back at the bicontinuous scaffolding, we can observe how this structure helps to explain this curling phenomenon which occurs with yarn relaxation. When observing the curling at the side edges of the knit, we can consider how each stitch, comprised of one left handed helicoid and one right handed helicoid, produce a balanced pair. This stitch is then balanced out on either side by another opposing helicoid from the next stitch. However, at each edge, we are left with a stitch without an opposing helicoid to balance it out, producing curling behaviour which follows the direction of the helicoid at each edge. (Figure 8.)

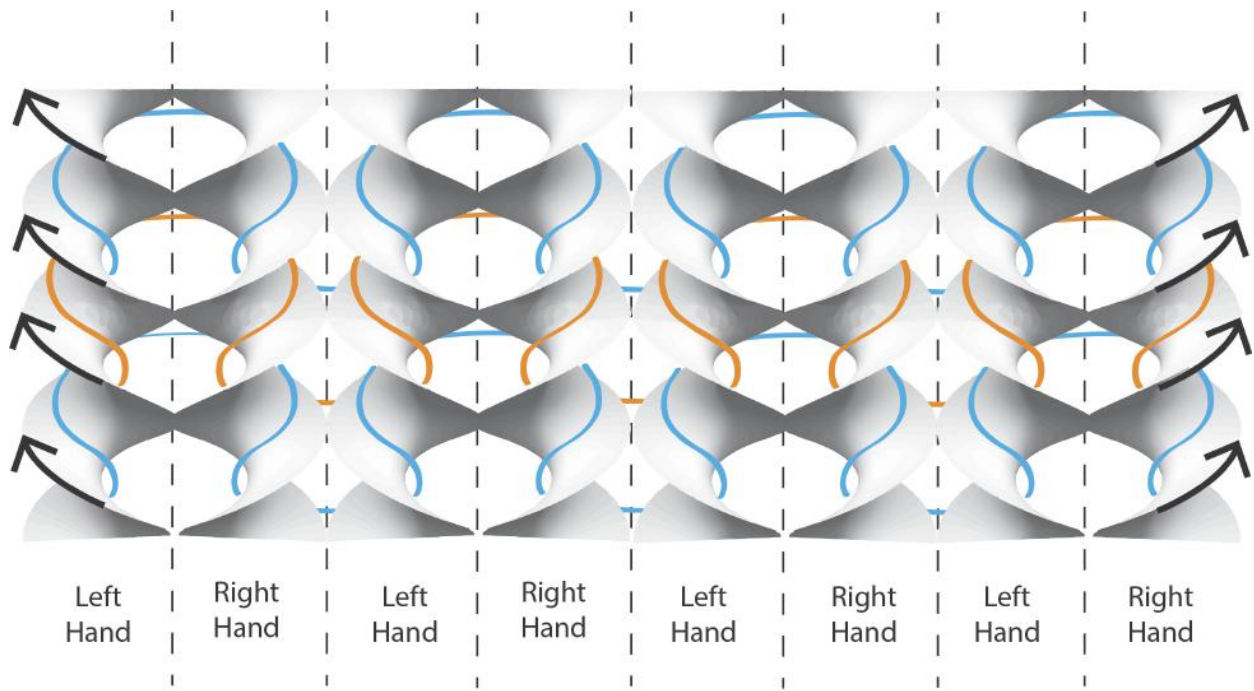

Figure 8. Direction of side edge curling in weft knits as shown by the helicoid lattice.

By considering the relationship of the yarn path to the helicoid geometry we can also explain the curling which occurs on the top and bottom edges of the knit. Following the yarn path at the topmost edge, the yarn for each loop moves around the centre of a LR helicoid pair. On the bottom, the yarn between each loop moves around the centre of a RL helicoid pair. When we compare the geometries of the LR and RL helicoid pairs, we can observe that the LR pair has a hole through the centre with an axis which moves down and back through the plane of the fabric. In the hole in the RL pair, this axis moves up and back through the plane. (Figure 9.a) When viewed from a side perspective, with these central axes depicted, we can visualise how the yarn of the loop structure pulls down on the helicoid pair at the top edge, and up on the helicoid pair at the bottom edge, initiating the curling of both edges towards the centre of the fabric. (Figure 9.b)

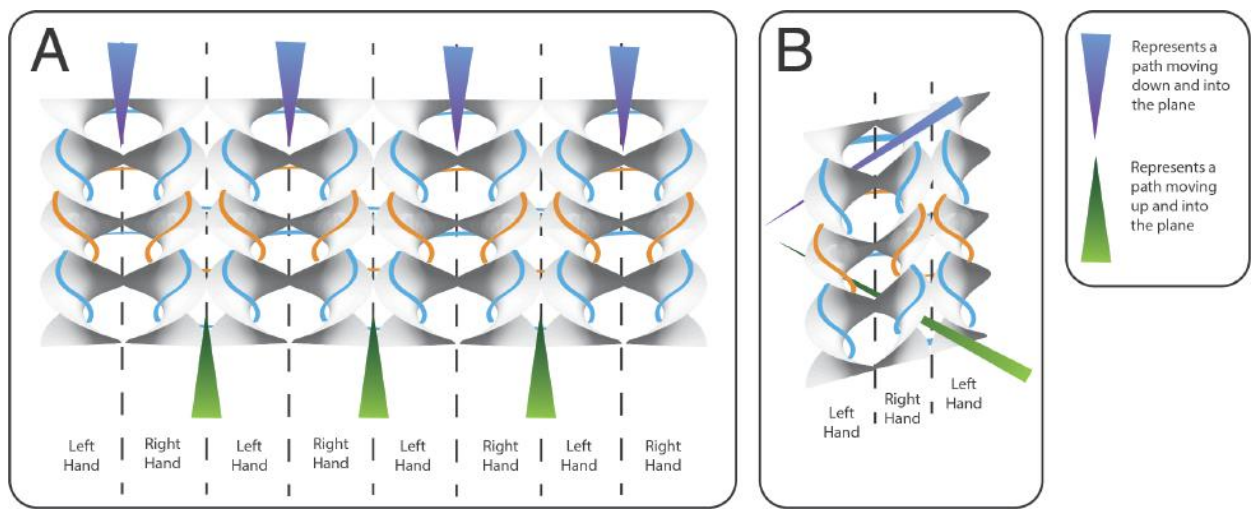

Figure 9. Diagram of yarn path for knit loop structure on the helicoid scaffold, with central axes depicted.

By changing the underlying lattice of helicoids we can create a variety of bicontinuous surfaces. This same concept could be applied to dynamic modelling of other textile structures, such as warp knit, woven and crocheted fabric, as well as for development of novel textile architectures.

5. Conclusion

Through a combination of artistic, scientific, and design-based thinking, we have developed a novel method of mathematically modelling yarn pathways in textile architectures using families of bicontinuous surfaces. Our physical prototype demonstrates how such a parametric model could help designers and engineers of textile, be it makers or manufacturers, better predict textile relaxation behaviours prior to production. By recreating this system of dynamic modelling in the virtual realm, this model could provide the flexibility needed to support more customisable solutions for functional fabrics, as well as facilitate development of teaching tools for academic programs in textiles, to help drive innovation in the field.

6. Acknowledgements

C.E.K was supported by the National Science Foundation Graduate Research Fellowship under Grant No. DGE-10028090/DGE-1104459 and C.E.K and G.D were supported by the National Science Foundation (CMMI #1537720). Any opinion, findings, and conclusions or recommendations expressed in this material are those of the authors(s) and do not necessarily reflect the views of the National Science Foundation. M.T. and R.D.K. were supported by a Simons Investigator Grant from the Simons Foundation to R.D.K. The authors would also like thank Christina Kara, the lab manager of the Shima Seiki Haute Tech Lab and Robert LehRich of Shima Seiki USA.

7. Declaration of Interests

The authors have no conflicts of interest to declare.